\documentclass[pra,twocolumn,aps,showpacs]{revtex4}
\usepackage{epsfig}
\usepackage{bm}
\begin{document}
\title{Mass imbalance effect in resonant Bose-Fermi mixtures}
\author{Elisa Fratini and Pierbiagio Pieri}
\affiliation{School of Science and Technology, Physics Division, 
University of Camerino and CNISM, Via Madonna delle Carceri 9, I-62032 Camerino, Italy}
\date{\today}

\begin{abstract}
We consider a homogeneous Bose-Fermi mixture, with the boson-fermion interaction tuned by a Fano-Feshbach resonance, in the presence of mass and density imbalance between the
two species.
By using many-body diagrammatic methods, we first study the finite-temperature phase diagram for the specific case of the mass-imbalanced mixture $^{87}$Rb $-^{40}$K for different values of the density imbalance.
We then analyse the quantum phase transition associated with the disappearance  at zero temperature of the boson condensate above a critical boson-fermion coupling.  
We find a pronounced dependence of the critical coupling on the mass ratio and a weak dependence on the density imbalance. For a vanishingly small boson 
density, we derive, within our approximation, the asymptotic expressions for the critical coupling in the limits of small and large mass ratios. These expressions are relevant also for the polaron-molecule transition in a Fermi mixture at small and large mass ratios.
The analysis of the momentum distribution functions
at sufficiently large density imbalances shows an interesting effect in the bosonic momentum distribution due to the simultaneous presence of composite fermions and unpaired
fermions.
\end{abstract}

\pacs{03.75.Ss,03.75.Hh,32.30.Bv,74.20.-z}
\maketitle

\section{Introduction}\label{intro}

An impressive series of experiments has been realized with ultracold atomic gases over the last years, reproducing systems, or physical situations, 
relevant to several areas of physics. 
The use of Fano-Feshbach resonances to control the interaction strength between particles, in particular, has been the cornerstone of many of these recent experimental 
achievements.

Besides providing analog quantum models and simulators for systems of interest in different fields of physics, ultracold gases offer also the possibility to construct 
novel many-body systems with no corresponding counterparts in other domains of physics. 
Resonant Bose-Fermi mixtures constitute an interesting example in this respect, and have been the object of active theoretical and experimental investigation recently.

Nonresonant Bose-Fermi mixtures were initially studied theoretically and experimentally in~\cite{Viv00,Yi01,Alb02,Viv02,Rot02,Li03,Lew04} and \cite{Mod02}, respectively.
Bose-Fermi mixtures in the presence of a  {\em narrow}
Fano-Feshbach resonance were then considered in the theoretical works~\cite{Dic05,Pow05,Avd06,Bor08,Mar08}. For a narrow resonance one has to take into account explicitly the molecular state forming 
in the closed channel, leading to a Hamiltonian which includes three different species (bosons, fermions, and molecules) from the outset. 

Both cases of narrow or broad resonance are, however, relevant to current experiments in Bose-Fermi mixtures~\cite{Gun06,Osp06,Osp06b,Zir08,Ni08,Wu11,Park11,Heo12}, depending on the mixture
and/or the resonance actually chosen in the experiment.
A broad Fano-Feshbach resonance is characterized by the smallness of the effective range parameter $r_0$ of the boson-fermion scattering amplitude 
 with respect to both the average interparticle distance and the boson-fermion scattering length $a$~\cite{Sim05}.
Under these conditions, the system can be described by a Hamiltonian made just by bosons and fermions mutually interacting via an attractive contact potential.

Initial works studying Bose-Fermi mixtures in the presence of a broad resonance focused on lattice models~\cite{Kag04,Bar08,Pol08,Riz08,Tit09,Mar09}, 
or considered separable interactions, as inspired by nuclear-physics models~\cite{Sto05}.
The continuum case in the presence of an attractive contact potential was first tackled in Ref.~\cite{Wat08}, which concentrated on the thermodynamic properties of the condensed phase
of a Bose-Fermi mixture at zero temperature.

A first study of the competition between Bose-Fermi pairing and boson condensation in a broadly resonant Bose-Fermi mixture has been presented in a previous work by us \cite{Fra10}. 
By using many-body diagrammatic theory, with a $T$-matrix approximation for the bosonic and fermionic self-energies, we were able to show that, for increasing Bose-Fermi attraction, 
the boson-fermion pairing correlations progressively reduce the boson condensation temperature and make 
it eventually vanish at a critical coupling above which the condensate is completely depleted (thus revealing the presence of a quantum phase transition at zero temperature).
In the present paper we extend the work of Ref.~\cite{Fra10} by analyzing the effect of a mass imbalance between the bosonic and fermionic species.
We present results both at finite temperature (for a specific mass ratio) and at zero temperature (for several mass ratios). The effect of mass imbalance in a broadly resonant Bose-Fermi 
mixture has been studied recently also in  Ref.~\cite{Lud11}, within a path-integral approach which is complementary to our approach  and that was limited to zero temperature only.  
A comparison between our results and those of Ref.~\cite{Lud11} will be discussed later on in our paper.

The paper is organized as follows. In section \ref{formulation} we present the theoretical formalism and the  main equations describing the physical system of interest in our paper. 
Section \ref{fintemp} reports the numerical results at finite temperature for a $^{87}$Rb $-^{40}$K mixture. 
The phase diagram for different values of the density imbalance is presented, together with the curves for the boson and fermion chemical potentials at the critical temperature. 
Section \ref{qpt} treats the zero temperature limit. We report the results for the dependence of the critical coupling on the mass ratio and density imbalance, the corresponding chemical potentials, and the momentum distribution functions. Section \ref{conclusions} presents finally our conclusions.

\section{Formalism}\label{formulation}
We consider a homogeneous mixture, composed by single-component fermions and bosons. The boson-fermion interaction is assumed to be tuned by a broad Fano-Feshbach resonance, as in most of current experiments
 on Bose-Fermi mixtures~\cite{Gun06,Osp06,Osp06b,Zir08,Ni08,Wu11,Park11}. Under this condition, the boson-fermion interaction can be adequately described by an attractive point-contact potential.
We consider then the the following (grand-canonical) Hamiltonian:
\begin{eqnarray}
H&=&\sum_{s}\int\! d {\bf r} \psi^{\dagger}_s({\bf r})(-\frac{\nabla^2}{2 m_s}-\mu_s)
\psi_s({\bf r}) \nonumber\\
&+& v_0 \int\! d{\bf r} \psi^{\dagger}_{\rm B}({\bf r})\psi^{\dagger}_{\rm F}({\bf r})
\psi_{\rm F}({\bf r})\psi_{\rm B}({\bf r})
\label{hamiltonian}.
\end{eqnarray}
Here $\psi^{\dagger}_s({\bf r})$, creates a particle of mass $m_s$ and chemical potential $\mu_s$ at 
spatial position ${\bf r}$, where $s$=B,F indicates the boson and 
fermion atomic species, respectively, while $v_0$ is the bare strength
of the contact interaction (we set $\hbar=k_{\rm B}=1$ throughout this paper). 
 As for two-component Fermi gases~\cite{Pie00}, 
the ultraviolet divergences associated with the contact  interaction in (\ref{hamiltonian}) are eliminated
by expressing the bare interaction $v_0$ in terms of the boson-fermion scattering length $a$:
\begin{equation}
\frac{1}{v_0}=\frac{m_r}{2\pi a}-\int \! \frac{d{\bf k}}{(2 \pi)^3} 
\frac{2 m_r}{{\bf k}^2}\;,
\end{equation}
where $m_r=m_{\rm B} m_{\rm F}/(m_{\rm B}+m_{\rm F})$ is the reduced mass of the boson-fermion system. 

We have not considered in the Hamiltonian (\ref{hamiltonian}) an explicit boson-boson interaction term. In the physical systems relevant to experiments, the corresponding  nonresonant scattering length is normally small and positive. In the homogeneous and normal system we are going to consider, it yields then only a mean-field shift of the boson chemical potential. Note however that, according to the variational analysis of Ref.~\cite{Yu11}, the assumption of having a homogeneous system actually requires the boson-boson scattering length to exceed a certain value, in order to guarantee the mechanical stability of the Bose-Fermi mixture against collapse. We thus implicitly assume to work in this stable regime. Note finally that a Fermi-Fermi $s-$wave scattering length is excluded by Pauli principle.

A natural length scale for the system is provided by the average interparticle distance $n^{-1/3}$ (where $n=n_{\rm B}+n_{\rm F}$ is the total particle-number density,  $n_{\rm B}$ and $n_{\rm F}$ being the individual boson and fermion particle-number density, respectively). We thus introduce a fictitious Fermi momentum of the system $k_{\rm F}\equiv (3 \pi^2 n)^{1/3}$ (as for an equivalent two-component Fermi gas with a density equal to the total density of the system), and use the dimensionless coupling parameter $g=(k_{\rm F} a)^{-1}$ to describe the strength of the interaction. 
In the weak-coupling limit, where the scattering length $a$ is small and negative and $ g \ll-1$, the two components behave essentially as Bose and Fermi ideal gases. In the opposite, strong-coupling, limit where $a$ is small and positive and $g \gg 1$, the system is expected to be be described in terms of composite fermions, i.e. boson-fermion pairs with a binding energy $\epsilon_0=1/(2 m_r a^2)$, and excess fermions.
[We restrict our analysis to mixtures where the number of bosons never exceeds the number of fermions. This is because only in this case a quantum phase transition associated with the disappearance of the condensate is possible \cite{Fra10}. In addition, mixtures with $n_{\rm B}> n_{\rm F}$ are expected to be severely affected by three-atom losses.]

\begin{figure}[t]
\epsfxsize=8cm
\epsfbox{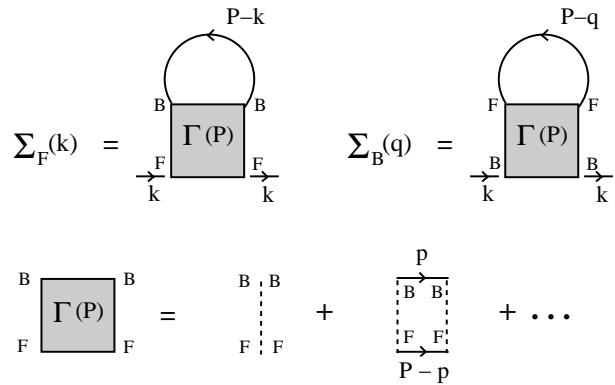}
\caption{(Color online) T-matrix diagrams for the fermionic and bosonic self-energies in the normal phase. Full lines represent bare bosonic (BB) and fermionic (FF) Green's functions. Broken lines represent bare boson-fermions interactions $v_0$.} 
\label{diagrams}
\end{figure}
The physical behavior of a Bose-Fermi mixture across all the resonance is captured by the $T$-matrix set of diagrams for the boson and fermion self-energies represented in Fig.\ref{diagrams}.
As shown in our previous work  \cite{Fra10}, this set of diagrams is able to recover the expected physical behavior in both the weak- and strong-coupling limits, thus providing a meaningful theoretical framework for the
whole resonance. The corresponding equations for the bosonic and fermionic self-energies $\Sigma_{\rm B}$ and $\Sigma_{\rm F}$ and many-body $T-$matrix $\Gamma$ read:
\begin{eqnarray}\label{selfb}
\Sigma_{\rm B}(q)&=&-T\int\!\!\frac{d {\bf P}}{(2\pi)^{3}}\sum_{m}G_{\rm F}^{0}(P-q)\Gamma(P)\\
\label{selff}
\Sigma_{\rm F}(k)&=& T\int\!\!\frac{d {\bf P}}{(2\pi)^{3}}\sum_{m}G_{\rm B}^{0}(P-k)\Gamma(P),
\end{eqnarray}
\begin{eqnarray}
&&\Gamma({\bf P}, \Omega_m)=- \left\{\frac{m_{r}}{2\pi a}+\int\!\!\frac{d{\bf p}}{(2\pi)^{3}}
\right.\nonumber\\
&&\times \left.\left[\frac{1-f[\xi_{\rm F}({\bf P}-{\bf p})]+b[\xi_{\rm B}({\bf p})]}{\xi_{\rm F}({\bf P}-{\bf p})+\xi_{\rm B}\left({\bf p}\right)-i\Omega_{m}}
-\frac{2m_{r}}{{\bf p}^{2}} \right]\right\}^{-1}.
\label{gamma}
\end{eqnarray}
Here $q=({\bf q},\omega_{\nu})$, $k=({\bf k},\omega_{n})$,  $P=({\bf P},\Omega_{m})$, where $\omega_{\nu}=2\pi\nu T$  and $\omega_{n}=(2n+1)\pi T$, $\Omega_m=(2m+1)\pi T$ 
are bosonic and fermionic Matsubara frequencies, respectively, ($\nu,n,m$ being integer numbers), while $f(x)$ and $b(x)$ are the Fermi and Bose distribution functions at temperature $T$, and
$\xi_s({\bf p})={\bf p}^2/(2m_s)-\mu_s$. 

The self-energies (\ref{selfb}) and (\ref{selff}) determine the dressed Green's functions $G_s$ via the Dyson's equation $G_s^{-1}=G_s^{0 \; -1}-\Sigma_s$ (with the bare Green's functions given by $G_{\rm B}^0(q)=[i\omega_{\nu}-\xi_{\rm B}({\bf q})]^{-1}$ and 
$G_{\rm F}^0(k)=[i\omega_n-\xi_{\rm F}({\bf k})]^{-1}$). 
The dressed Green's functions $G_s$ allow to calculate the boson and fermion momentum distribution functions through the equations:
\begin{eqnarray}\label{nbq}
n_{\rm B}({\bf q})&=&- T \sum_{\nu}G_{\rm B}({\bf q},\omega_{\nu})\,e^{i\omega_{\nu} 0^+}\\
\label{nfk}
n_{\rm F}({\bf k})&=& T \sum_{n}G_{\rm F}({\bf k},\omega_{n})\,e^{i\omega_{n} 0^+}\,,
\end{eqnarray}
which in turn determine the boson and fermion number densities:
\begin{eqnarray}\label{nb}
n_{\rm B}&=&\int\!\!\frac{d {\bf q}}{(2\pi)^{3}} n_{\rm B}({\bf q})\\
\label{nf}
n_{\rm F}&=&\int\!\!\frac{d {\bf k}}{(2\pi)^{3}} n_{\rm F}({\bf k}).
\end{eqnarray}
At fixed densities and temperature, the two coupled equations (\ref{nb}) and (\ref{nf}) fully determine the boson and fermion chemical potentials, and therefore the thermodynamic properties of the Bose-Fermi mixture in the normal phase.

Coming from the normal phase, the condensation of bosons starts when
the condition
\begin{equation}\label{eqTc}
 \mu_{\rm B}-\Sigma_{\rm B}(q=0)=0
\end{equation}
is first met. Eq.~(\ref{eqTc}) corresponds to the requirement of a divergent occupancy of the boson  momentum 
distribution at zero momentum: $\lim_{\bf q\to 0} n_{\rm B}({\bf q})=\infty$. \cite{popov}

\subsection{The zero-temperature limit}\label{zerotemp}

We pass now to consider the zero-temperature limit of the previous finite-temperature formalism. When $T\to 0$, the spacing $2\pi T$ between two consecutive Matsubara frequencies tends to zero. The sums over discrete Matsubara frequencies can then be replaced by integrals over continuous frequencies, provided the corresponding integrands are not too singular~\cite{footnote-zeroT}.  
We have then the equations 
\begin{eqnarray}\label{selfbzero}
\Sigma_{\rm B}(q)&=&-\int\!\!\!\!\frac{d {\bf P}}{(2\pi)^{3}}\int\!\!\frac{d {\Omega}}{2\pi}G_{\rm F}^{0}(P-q)\Gamma(P)\\
\label{selffzero}
\Sigma_{\rm F}(k)&=&\int\!\!\!\!\frac{d {\bf P}}{(2\pi)^{3}}\int\!\!\frac{d {\Omega}}{2\pi}G_{\rm B}^{0}(P-k)\Gamma(P),
\end{eqnarray}
for the bosonic and fermionic self-energies, where $q=({\bf q}, \omega_{\rm B})$, $k=({\bf k}, \omega_{\rm F})$, $P=(\textbf{P}, \Omega)$ while the frequencies $\omega_{\rm B}$, $\omega_{\rm F}$ and $\Omega$ are now continuous variables.

Similarly, the equations (\ref{nbq}) and (\ref{nfk}) for the momentum distribution functions are changed to
\begin{eqnarray}\label{nbzero}
n_{\rm B}({\bf q})&=&-\int\!\!\frac{d {\omega_{\rm B}}}{2\pi}G_{\rm B}({\bf q},\omega_{\rm B})e^{i\omega_{\rm B} 0^+}\\
\label{nfzero}
n_{\rm F}({\bf k})&=&\int\!\!\frac{d {\omega_{\rm F}}}{2\pi}G_{\rm F}({\bf k},\omega_{\rm F})e^{i\omega_{\rm F} 0^+},
\end{eqnarray}
from which the number densities can be calculated as before. 

A closed-form expression can be finally derived for the many-body T-matrix $\Gamma(P)$ at zero temperature, when the Fermi and Bose distribution functions appearing in Eq.~(\ref{gamma}) are replaced by step functions. We report in particular the expression for $\Gamma(P)$ when $\mu_{\rm F} > 0$ and $\mu_{\rm B} < 0$, which is in practice the only one relevant to our calculations at zero temperature.

We have
\begin{equation}
\Gamma(P)=- \left[ \frac{m_{r}}{2\pi a}-\frac{m_r^{\frac{3}{2}}}{\sqrt{2}\pi}\sqrt{\frac{{\bf P}^2}{2M}-2\mu-i\Omega}-I_{\rm F}(P)\right]^{-1},
\label{gammazero}
\end{equation}
where we have defined $M=m_{\rm B}+m_{\rm F}$ and $\mu=(\mu_{\rm B}+\mu_{\rm F})/2$, while $I_{\rm F}(P)$, which results from the integration of the term with the Fermi function in  Eq.~(\ref{gamma}), is given by
\begin{eqnarray}
I_{\rm F}(P)&=&\frac{m_{\rm B} \left(k_{\mu_{\rm F}}^2-k_{{\bf P}}^2-k^2_\Omega\right)}{8 \pi^2 |{\bf P}|}
\ln\left[ \frac{(k_{\mu_{\rm F}}+k_{{\bf P}})^2-k_\Omega^2}{(k_{\mu_{\rm F}}-k_{{\bf P}})^2-k_\Omega^2}\right]
\nonumber\\
&-& \frac{m_r k_\Omega}{4\pi^2}\left\{
 \ln\left[\frac{(k_{\mu_{\rm F}}+k_\Omega)^2-k_{{\bf P}}^2}{k_{{\bf P}}^2-(k_{\mu_{\rm F}}-k_\Omega)^2}\right]-i \pi {\rm sgn}(\Omega)\right\}\nonumber\\
&+&\frac{m_r k_{\mu_{\rm F}}}{2\pi^2},\label{If}
\end{eqnarray}
where $k_{\mu_{\rm F}}\equiv\sqrt{2 m_{\rm F}\mu_{\rm F}}$, $k_{{\bf P}} \equiv \frac{m_{\rm F}}{M} P$, while 
\begin{equation}
k_\Omega \equiv (2 m_r)^{\frac{1}{2}}\sqrt{-\frac{{\bf P}^2}{2M}+2\mu+i\Omega}.
\end{equation}

Equations (\ref{selfbzero})-(\ref{If}) determine the thermodynamic properties of a Bose-Fermi mixture at zero temperature in the absence of boson condensation. 
They are thus relevant for a sufficiently strong coupling $g$, such that the system remains in the normal phase even at zero temperature.
In particular, upon lowering the coupling constant $g$, the condensation will start at a critical coupling $g_c$, when the condition (\ref{eqTc}) is first satisfied.

\section{Finite-temperature results}\label{fintemp}
The theoretical approach developed in section \ref{formulation}, can be used to explore the normal phase of a homogeneous Bose-Fermi mixture at arbitrary values of the boson and fermion masses and densities.
In this section, where  we present finite-temperature results, we focus on the specific mass ratio $m_{\rm B}/m_{\rm F}=87/40$, relevant for the  $^{87}$Rb $-^{40}$K mixture. More general mass ratios will be considered at zero temperature. 

\subsection{Critical temperature}
\begin{figure}[t]
\epsfxsize=8cm
\epsfbox{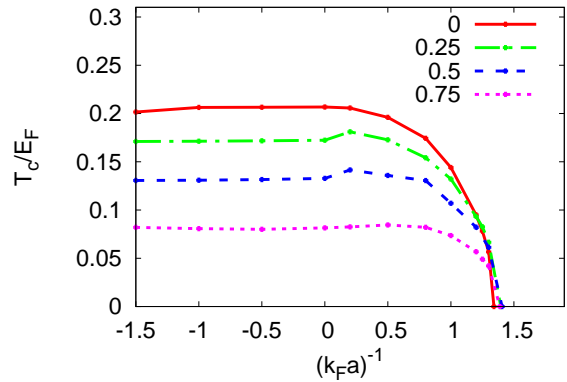}
\caption{(Color online) Critical temperature (in units of $E_{\rm F}=k_{\rm F}^2/ 2 m_{\rm F}$) for condensation of bosons as a function of 
the boson-fermion coupling  $(k_{\rm F} a)^{-1}$ for different values of 
the density imbalance $(n_{\rm F}-n_{\rm B})/(n_{\rm F}+n_{\rm B})$ in a mixture with $m_{\rm B}/m_{\rm F}=87/40$.} 
\label{Tc_8740}
\end{figure}

Figure \ref{Tc_8740} presents the dependence of the condensation critical temperature on the boson-fermion coupling $(k_{\rm F} a)^{-1}$  for a $^{87}$Rb $-^{40}$K mixture, at different values of the density imbalance $(n_{\rm F}-n_{\rm B})/(n_{\rm F}+n_{\rm B})$.
The critical temperature was obtained by solving numerically Eqs.~(\ref{selfb})-(\ref{nf}), supplemented by the condition (\ref{eqTc}), while the ending point at $T=0$ was calculated independently by solving the equations (\ref{selfbzero})- (\ref{If}) (with the condition (\ref{eqTc}) defining now the critical coupling $g_c$ at zero temperature).  The matching between finite-temperature and zero-temperature results confirms the validity of the equations derived in the zero-temperature limit, while providing simultaneously a check of the
numerical calculations.

The overall behavior of the critical temperature as a function of coupling 
is similar to what found for equal masses~\cite{Fra10}. The critical
temperature starts from the noninteracting value ($T_0= 3.31 n_{\rm B}^{2/3}/m_{\rm B}$) in the weak-coupling limit and eventually vanishes
at a critical coupling $g_c$,  when the effect of the boson-fermion coupling is so strong that the Bose-Einstein condensate is completely depleted, in favor of the formation of molecules.
 The weak dependence of the critical 
coupling on the density imbalance previously found for equal masses is confirmed also for this case with
different masses. In this case, the critical coupling varies in the range $1.3\div 1.4$, to be compared with the range $1.6\div 1.8$ found for $m_{\rm B}=m_{\rm F}$. 
 
A minor difference with respect to the case with equal masses is finally the presence of a weak maximum in the critical temperature, which reaches a value slightly above the noninteracting value $T_0$ before its final decrease. This feature is more pronounced for intermediate values of the density imbalance. We attribute this feature to the delicate balance between the effects of the boson-fermion interaction on the boson dispersion,  and the condensate depletion due to molecular correlations. The boson dispersion may in fact be hardened by the interaction (similarly to what one finds for a dilute repulsive Bose gas \cite{Holz03}), thus leading to an increase of the critical temperature. The predominance of one effect over the other one depends on a fine tuning of the mass ratio and density imbalance, and may explain the different behavior found for different values of these parameters.

\begin{figure}[t]
\epsfxsize=8cm
\epsfbox{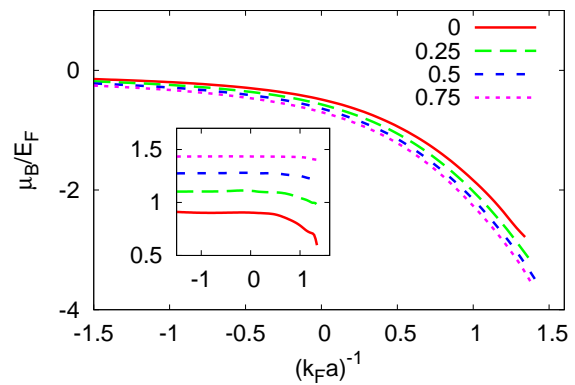}
\caption{(Color online) Bosonic chemical potential at the critical temperature $T_c$ as a 
function of the boson-fermion coupling $(k_{\rm F} a)^{-1}$ for different
values of the density imbalance $(n_{\rm F}-n_{\rm B})/(n_{\rm F}+n_{\rm B})$ in a mixture with $m_{\rm B}/m_{\rm F}=87/40$. The 
corresponding fermionic chemical potential is reported in the inset.} 
\label{muTc_8740}
\end{figure}

\subsection{Chemical potentials}
 Figure \ref{muTc_8740} reports the coupling dependence of the chemical potentials $\mu_{\rm B}$ and $\mu_{\rm F}$  at the critical temperature for a $^{87}$Rb $-^{40}$K  mixture at the same values of the density imbalance considered in 
Fig.~\ref{Tc_8740}. 

 In this case, the critical boson and fermion chemical potentials present the same qualitative behavior already found for a mixture with equal masses~\cite{Fra10}. 
The boson chemical potential decreases markedly with increasing coupling,  and changes from $\mu_{\rm B}\approx 2 \pi  n_{\rm F} a/m_r$ for weak coupling to $\mu_{\rm B}\approx -\epsilon_0$ for strong coupling, with a small dependence on the density imbalance.

The fermion chemical potential (reported in the inset)  remains instead almost constant across the whole resonance. The decrease of the chemical potential due to the attractive interaction with the bosons is, in fact, partially compensated by the decreasing of the temperature when moving along the critical line (which increases $\mu_{\rm F}$) and by the Pauli repulsion between unpaired fermions and Bose-Fermi pairs.

\section{Zero-temperature results}\label{qpt}

The behavior of the critical temperature as a function of the boson-fermion coupling discussed in the previous Section, evidenced the presence of a quantum phase transition at zero temperature associated with a transition between
a superfluid phase with a boson condensate to a normal phase, where the condensate is completely depleted.
In this Section we examine in more detail this quantum phase transition by solving numerically the equations formulated at exactly zero temperature. 

\subsection{Critical couplings and chemical potentials}

\begin{figure}[t]
\epsfxsize=8cm
\epsfbox{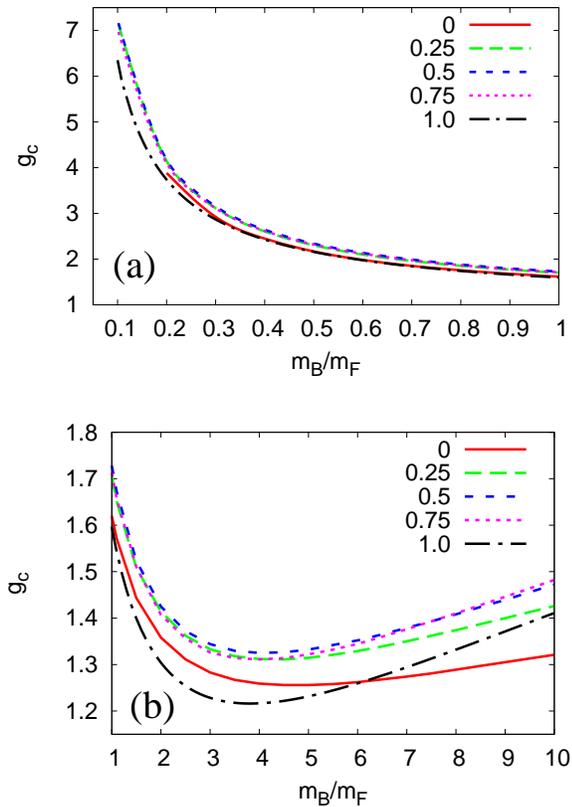}
\caption{(Color online) Critical coupling $g_c$ as a function of the mass ratio ($m_{\rm B}/m_{\rm F}$) for different values of the density imbalance $(n_{\rm F}-n_{\rm B})/(n_{\rm F}+n_{\rm B})$. Panels (a) and (b) correspond to the ranges $m_{\rm B}/m_{\rm F} \le 1$ and $m_{\rm B}/m_{\rm F} \ge 1$, respectively.} 
\label{gc_massb}
\end{figure}

Figure \ref{gc_massb} reports the critical coupling $g_c$ as a function of the mass ratio $m_{\rm B}/m_{\rm F}$, with two distinct panels for the cases  $m_{\rm B}/m_{\rm F} \le 1$ and $m_{\rm B}/m_{\rm F} \ge 1$.
The different curves reported in Fig.~\ref{gc_massb} correspond to different values of the density imbalance,  ranging from the density-balanced case to the fully imbalanced one $(n_{\rm F}-n_{\rm B})/(n_{\rm F}+n_{\rm B})=1.0$, which represents the system with just one boson immersed in a Fermi sea.
 This  is actually the same as a spin-down fermion sorrounded by a Fermi sea of spin-up fermions, since
for a single particle the statistics is irrelevant.  
The critical coupling reduces thus to that for the polaron-to-molecule transition, 
recently studied in the context of strongly imbalanced two-component Fermi gases~\cite{pro08,vei08,mas08,mor09,pun09,com09,sch09}.
The equations governing the one boson limit are reported in the Appendix.

The critical coupling is strongly influenced by the mass ratio, especially for $m_{\rm B}/m_{\rm F}<1$. In this case, for all values of the density imbalance, $g_c$ increases very rapidily as $m_{\rm B}/m_{\rm F}$ is decreased.
For the single boson problem, an asymptotic expansion of the equations determining $g_c$ for $m_{\rm B}/m_{\rm F}\to 0$ yields the result
\begin{eqnarray}
 g_c&\approx&\frac{4 \cdot 2^{1/3}}{3\pi} \frac{m_{\rm F}}{m_{\rm B}}+ \frac{32 \cdot 2^{1/3}}{15\pi}\label{asysmall}\\
&\approx& 0.535 \frac{m_{\rm F}}{m_{\rm B}} +0.856
\end{eqnarray} 
which proves quite accurate even before the asympotic regime is reached (the deviation from the numerical solution is 15\% for $m_{\rm B}/m_{\rm F}=1$ and  2\% for $m_{\rm B}/m _{\rm F}=0.1$). 
[See the Appendix for the details of the derivation of Eq.~(\ref{asysmall}).]

The rapid increase of the critical coupling for the
polaron-to-molecule transition when $m_\downarrow \to 0$ was already noticed in Ref.~\onlinecite{com09}, even though no asymptotic expression was reported there. Note however that, according to the analysis of the polaron-to-molecule transition  of Ref.~\cite{Mat11}, for a mass ratio 
$m_{\downarrow}/m_{\uparrow}\lesssim 0.15$ the molecular state acquires a finite momentum in its ground state. Similar results were obtained in 
Ref.~\cite{Song11}.  
The equations for the single boson problem here adopted assume that the formation of the molecule occurs at zero center of mass momentum (as in Refs.~\onlinecite{pro08,vei08,mas08,mor09,pun09,com09,sch09}). The curve corresponding to the single boson problem in Fig.~\ref{gc_massb} 
may thus change  for $m_{\rm B}/m_{\rm F}\lesssim 0.15$, after taking into account the possibility of pairing at finite momentum.
We expect however this change to be minor on the basis of our calculations with a finite boson density, which allow for pairing at finite momentum and yield results close to the single-boson curve also 
for $m_{\rm B}/m_{\rm F} < 0.15$.

Note further that the study of the three body system with two equal fermions with mass $m_{\rm F}$ and one different particle with mass $m_{\rm B}$ interacting through a zero-range potential,  shows 
that for $m_{\rm B}/m_{\rm F}<0.0735=(13.607)^{-1}$ the system is
unstable due to a sequence of three-body bound states with energy $\to -\infty$.\cite{braaten06} A similar instability is expected to occur also in
 the many-body system with $N$ equal fermions plus one different particle 
(a recent work has proven indeed that the above critical value for the three-body system provides a lower bound for the location of the instability in the many-body system \cite{cor12}). 
We note, however, that the unboundness from below of the Efimov spectrum (and the associated global instability) occurs only for a pure zero-range interaction.
In a real system, physical two-body interactions will provide a natural cut-off at distances of the order of the van-der-Waals length $r_{{\rm vdW}}$, thus limiting
the position of the lowest Efimov level at an energy $\sim - 1/r_{{\rm vdW}}^2$. The global mechanical instability is thus avoided by the real system, even though the presence of Efimov states is expected to  
lead to an enhancement of three-body losses when they lie close to the three-particle or atom-dimer continuum. 
We further note in this context that the presence of the (non-universal) three-body bound-states recently found by Kartavtsev and Malykh \cite{Kar07} for  $(8.2)^{-1} > m_{\rm B}/m_{\rm F}>(13.607)^{-1}$ could also lead to enhanced three-body losses in this mass-ratio range.

We observe in any case that since our calculations were taken for  $m_{\rm B}/m_{\rm F}\ge 0.1$ ($m_{\rm B}/m_{\rm F} \ge 0.2$ for equal densities, due to numerical difficulties), the above three-body effects (which are out of the scope of the present theory) should affect our study only marginally.

Consistently with our previous results, we observe a weak dependence of $g_c$ on the density imbalance. Such a weak dependence on the densities remains valid also for $m_{\rm B} > m_{\rm F}$. 
In this case, all curves reach a minimum value of $g_c$ (=1.2$\div$1.3) for mass ratios $m_{\rm B}/ m_{\rm F}$ in the range 3.5$\div$5, after which they increase slowly with the mass ratio. 
For the single boson problem with a large mass we have obtained  the asymptotic expression (see Appendix): 
\begin{equation}
g_c\approx A(m_{\rm B}/m_{\rm F})-\frac{2^{2/3}}{A(m_{\rm B}/m_{\rm F})}
+ \frac{16}{3\pi}\frac{1}{A(m_{\rm B}/ m_{\rm F})^2}
\label{asylarge1} 
\end{equation}
where
\begin{equation}
A(m_{\rm B}/m_{\rm F})=\frac{2^{4/3}}{\pi}\left(\ln\frac{4 m_{\rm B}}{m_{\rm F}}-2\right) .
\label{asylarge2}
\end{equation}
One can see from Eqs.~(\ref{asylarge1}) and (\ref{asylarge2}) that at large mass ratios $g_c$ increases very slowly, with a logarithmic dependence on the mass ratios. Due to this log-dependence the leading behavior, 
$g_c\approx(2^{4/3}/\pi)\ln\frac{4 m_{\rm B}}{m_{\rm F}}$, is reached only at extremely large mass ratios. Equation~(\ref{asylarge1}), which includes the first two corrections to the leading behavior, 
provides a better approximation, the deviation from the numerical solution being 15\% for $m_{\rm B}/m_{\rm F}= 20$ and 1.5\% for $m_{\rm B}/m_{\rm F}=100$.

The increasing behavior of $g_c$ at both small and large mass ratios implies the existence of a minimum in the curve for $g_c$ at intermediate mass ratios, consistently with the results of Fig.~\ref{gc_massb}. Note however 
that in earlier work for the polaron-molecule transition, the critical coupling for the transition was found to move away from the BEC limit for increasingly heavier impurities. In particular, Ref.~\cite{com09} reported
that the critical coupling should approach the unitary limit for an infinitely heavy impurity. 
Similar results were also found in two-dimensions by M. Parish~\cite{par11} 
(with the critical coupling approaching the weak-coupling limit in this case). 
A reason for such a difference may be that our theory in the limit of a single boson reduces to the ``first level approximation'' of Ref.~\cite{com09}, with 
no particle-hole dressing of the molecule (while the polaron is described with the same accuracy obtained with the variational wave-function introduced by Chevy \cite{chevy}, which is deemed quite accurate for the polaron energy~\cite{com09}). The ``second level approximation'' of Ref.~\cite{com09} includes instead a particle-hole dressing of the molecule. This inclusion is  sufficient to recover in the strong-coupling limit the correct dimer-fermion scattering length, and shifts the position of the critical coupling  for equal masses from the value $(k_{{\rm F}} a)^{-1}= 1.27 c (=1.60)$ to the value $0.88 c (=1.11)$ (the factor $c=2^{1/3}$ appears here because of a different definition of $k_{\rm F}$  between us and the above references). The eventual (slow) increase of $g_c$ at large mass ratios could then be an artifact of the ``first order approximation''. Calculations at large mass ratios with alternative methods (such as fixed-node or diagrammatic Quantum Monte-Carlo methods) 
could definitively clarify this issue.

At equal densities, our results for $g_c$ as a function of the mass ratio agree well with the results  reported in Ref.~\cite{Lud11} for the same case. This is because the equations used in 
Ref.~\cite{Lud11} for calculating $g_c$ correspond, in a diagrammatic formalism, to the same choice of the self-energy as ours, but with the Dyson's equation expanded:
$G=G_0 + G_0\Sigma G_0$, instead of $G^{-1}=G_0^{-1}-\Sigma$. Even though such an expansion is justified only when $\Sigma$ is small , apparently it leads only to minor differences in the values for $g_c$. 
For instance, for equal masses and densities we obtained $g_c=1.62$, to be compared with $g_c=1.66$ in Ref.~\cite{Lud11}.
 
\begin{figure}[t]
\epsfxsize=8cm
\epsfbox{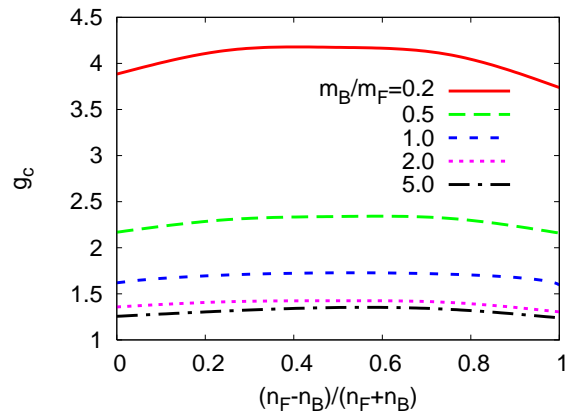}
\caption{(Color online) Critical coupling $g_c$ as a function of the density imbalance $(n_{\rm F}-n_{\rm B})/(n_{\rm F}+n_{\rm B})$, for different values of the mass ratio $m_{\rm B}/m_{\rm F}$.} 
\label{gc_imbal}
\end{figure}

According to the analysis of  Ref.~\cite{Lud11}, however,
only for sufficiently large values of the boson-boson scattering length $a_{\rm B}$, the critical coupling $g_c$ lies in a stable region of the phase diagram. For small values of $a_{\rm B}$, 
the second-order quantum phase transition between a condensed phase and a normal phase is in fact superseded by a phase separation between the two phases. 
According to our calculations, the compressibility matrix $\partial n_s/\partial \mu_{s'}$ is always positive in the normal phase, indicating that the second order phase transition here explored lies, at worst, in 
a metastable region of the phase diagram. In order to examine its absolute stability within our approach,  one should extend our study to the superfluid phase and make a comparison of the free energies for the normal 
and superfluid phases. This nontrivial extension is postponed to future work. We observe however, that  if ratios $a_{\rm B}/a$ of the order of 0.2-0.3 are sufficient to suppress phase separation in most of the phase 
diagram (as the results of Ref.~\cite{Lud11} seem to indicate), then the effect of $a_{\rm B}$ on $g_c$ will be minor.

The weak dependence of the critical coupling on the density imbalance is emphasized in  Fig.~\ref{gc_imbal}, which presents $g_c$ as a function of the density imbalance for some representative values of the mass ratio 
$m_{\rm B}/m_{\rm F}$.
All curves show a weak dependence on the density imbalance, with a weak  maximum at an intermediate imbalance (=0.55 for equal masses, and similarly for the other mass ratios considered here).

\begin{figure}[t]
\epsfxsize=8cm
\epsfbox{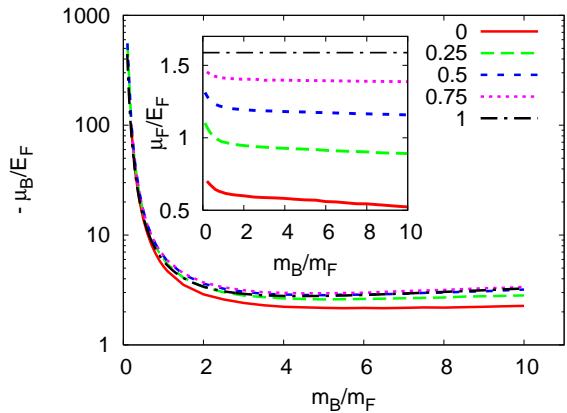}
\caption{(Color online) Bosonic chemical potential at the critical coupling $g_c$ as a function of the mass ratio $m_{\rm B}/m_{\rm F}$ for different values of the density imbalance 
$(n_{\rm F}-n_{\rm B})/(n_{\rm F}+n_{\rm B})$. The corresponding fermionic chemical potential is reported in the inset.}
\label{chem_potb}
\end{figure}

Figure~\ref{chem_potb} reports the chemical potentials at $g_c$ as a function of the mass ratio $m_{\rm B}/m_{\rm F}$, for different density imbalances. 
As it can be seen from the main panel, the (negative) bosonic chemical potential increases very rapidly  in absolute value for $m_{\rm B}/m_{\rm F}\to 0$.
This is because at $g_c$ the bosonic chemical potential is already close to its strong-coupling limit, $\mu_{\rm B}\approx -\epsilon_0$, such that the dimensionless ratio $|\mu_{\rm B}|/E_{\rm F}\approx 2 g^2 m_{\rm F}/m_r$.
Since $g_c \sim m_F/m_B$ for $m_B\to 0$, we have  $|\mu_{\rm B}|/E_{\rm F}\sim  (m_{\rm F}/ m_{\rm B})^3$ in this limit.

In the opposite limit of large $m_{\rm B}/m_{\rm F}$, the ratio $m_{\rm F}/m_r$ slowly increases and eventually saturates to 1 for large $m_{\rm B}$, such that $|\mu_{\rm B}|/E_{\rm F}$ follows the slow logarithmic increase of $g_c$ in this limit.

The fermionic chemical potential (reported in the inset) depends weakly on the mass imbalance, reflecting the weak dependence on $g_c$. As a matter of fact, the fermion chemical potential is determined essentially by the fermion density $n_{\rm F}$, independently from the coupling value or mass ratio.
 
\begin{figure}[t]
\begin{center}
\epsfxsize=8cm
\epsfbox{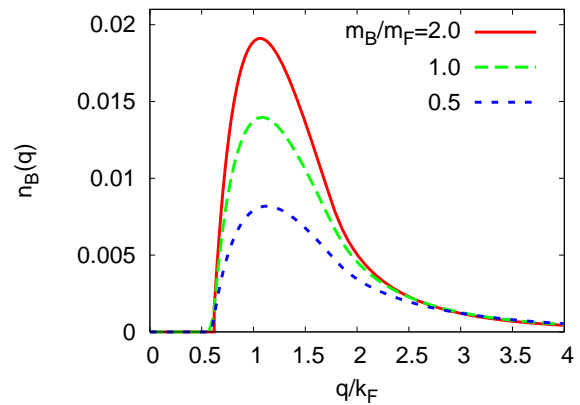}
\caption{(Color online) Bosonic momentum distribution curves at $g_c$  for a  fixed density imbalance  $(n_{\rm F}-n_{\rm B})/(n_{\rm F}+n_{\rm B})=0.75$ and different values of the mass ratio ($m_{\rm B}/m_{\rm F}$).}
\label{mom_distrb}
\end{center}
\end{figure}

\begin{figure}[t]
\epsfxsize=8cm
\epsfbox{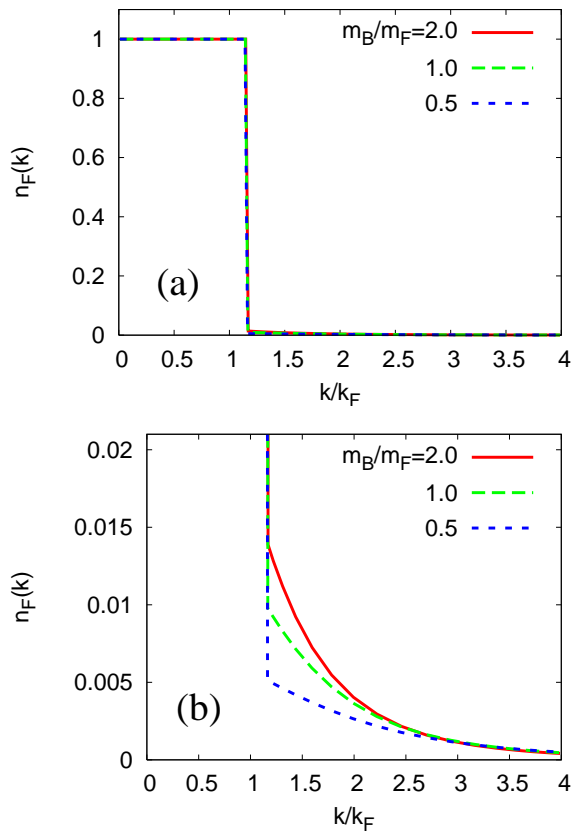}
\caption{(Color online) Fermionic momentum distribution curves at $g_c$ for a fixed density imbalance  $(n_{\rm F}-n_{\rm B})/(n_{\rm F}+n_{\rm B})=0.75$ and different values of the mass ratio $m_{\rm B}/m_{\rm F}$. The two panels correspond to a different choice of the vertical scale.}
\label{mom_distrf}
\end{figure}

\subsection{Momentum distribution functions}
 We pass now to study the momentum distribution functions  $n_{\rm B}(|{\bf q}|)$  and $n_{\rm F}(|{\bf k}|)$  for the  bosons and fermions, as obtained from Eqs.~(\ref{nbzero}) and (\ref{nfzero}), respectively.
We present results for a density imbalance   $(n_{\rm F}-n_{\rm B})/(n_{\rm F}+n_{\rm B})=0.75$, as to emphasize an interesting behavior of the bosonic momentum distribution function, which occurs only for a sufficiently large imbalance.
 
One can see, indeed, from Fig.~\ref{mom_distrb} that the bosonic momentum distribution function vanishes identically at low momenta.
This empty region extends from $|{\bf q}|=0$ up to a certain value 
$|{\bf q}|= q_0$, which is determined essentially only by the density imbalance (for the specific case of Fig.~\ref{mom_distrb}, $q_{0}\simeq 0.55$). 

The presence of the empty region can be interpreted as follows. For 
$n_{\rm F}\gg n_{\rm B}$, most of fermions remain unpaired and fill a Fermi sphere of radius $k_{\rm UF}\simeq[(n_{\rm F} - n_{\rm B})/6\pi^2]^{1/3}$, as Fig.~\ref{mom_distrf} for the fermionic distribution clearly shows. At $g_c$ and larger couplings, the bosons are instead bound into molecules that, being composite fermions, fill a Fermi sphere with a radius $P_{\rm CF}\simeq(n_{\rm B}/6\pi^2)^{1/3}$.  
Now, as the region $|{\bf k}| < k_{\rm UF}$ 
is already occupied by the unpaired fermions, only fermions with $|{\bf k}| > k_{\rm UF}$ 
participate to the molecule. Since the momentum of the molecule is given by the sum ${\bf P}={\bf q} + {\bf k}$, the constraints $|{\bf P}| < P_{\rm CF}$ and $|{\bf k}| > k_{\rm UF}$ then imply that only
 bosons with $|{\bf q}| > q_0= k_{\rm UF} - P_{\rm CF}$ participate to the molecule, leaving thus empty the region $|{\bf q}| < q_0$.
 
We have verified that the equation $q_0= k_{\rm UF} - P_{\rm CF}$ reproduces rather accurately the values of $q_0$ obtained numerically. In particular, the empty region does not exist at 
small density imbalance, when $k_{\rm UF} < P_{\rm CF}$. Note also that the initial rise of $n_{\rm B}(|{\bf q}|)$ after the threshold $q_0$ is due to the progressive increase of the 
phase-space volume corresponding to the ${\bf q}$'s satisfying the above constraints at a given ${\bf k}$. The saturation volume in phase-space is reached  for 
 $|{\bf q}|$ of the order of $k_{\rm UF}$, after which $n_{\rm B}(|{\bf q}|)$ starts to decrease, following eventually at sufficiently large wave-vectors a  molecular-like internal 
wave-function $n_{\rm B}(|{\bf q}|)\simeq n_{\rm M} |\phi(|{\bf q}|)|^2 $, 
where  $|\phi(|{\bf q}|)|^2$ can be approximated by the two-body 
normalized wave-function $\phi(|{\bf q}|)=(8\pi a^3)^{1/2}/({\bf q}^2 a^2+1)$, 
while the coefficient $n_{\rm M}\simeq n_{\rm B}$ can be interpreted as the molecular density. 

\begin{figure}[t]
\epsfxsize=8cm
\epsfbox{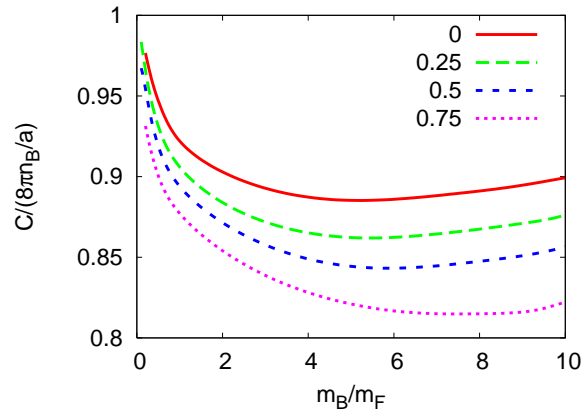}
\caption{Contact constant $C$, normalized to its strong-coupling limit $8\pi n_{\rm B}/a$, at  $g_c$ vs.~the mass ratio $m_{\rm B}/m_{\rm F}$ for different values of the density imbalance  $(n_{\rm F}-n_{\rm B})/(n_{\rm F}+n_{\rm B})$.}
\label{contact}
\end{figure}

The above approximate expression for $n_{\rm B}(|{\bf q})$ 
accounts for the main difference in the curves calculated at different mass ratios (namely, the decreasing height of the curves when the mass ratio is lowered) due to the strong dependence of $g_c$, 
and then of $a$, on the mass ratio. 
Note further that the same kind of beaviour is seen in the fermionic distribution function at momenta $|{\bf k}| > k_{\rm UF}$, as it can be evinced from Fig.~\ref{mom_distrf} (b). The comparison
between   Fig.~\ref{mom_distrf} (b) and Fig.~\ref{mom_distrb} shows indeed that the bosonic  and fermionic distribution functions become identical as the momentum increases;
at large momenta $n_{\rm F}(|{\bf k}|) \sim n_{\rm B}(|{\bf k}|)\sim C/{\bf k}^4$, consistently with the universal large momenta behavior established in Ref.~\cite{tan}. In particular, within our approximation, one can prove 
by taking the large $|{\bf k}|$ limit in our expressions that the ``contact'' constant $C$ is given by 
$C= - 4 m_r^2 \int \frac {d^4 P}{(2 \pi)^4} \Gamma (P) e^{i \Omega 0^+}$. In the strong-coupling limit, where all bosons are bound into molecules, a comparison with the expression for the molecular internal wave-function then leads to the equation $C= 8\pi n_{\rm B}/a$. 
Figure~\ref{contact} reports  the contact constant $C$ at $g_c$ normalized to its strong-coupling limit value $8\pi n_{\rm B}/a$, as a function of the mass ratio $m_{\rm B}/m_{\rm F}$ for different values of the density imbalance. One can
see that the constant $C$ at $g_c$ is close to its strong-coupling limit expression for all cases considered, with the largest deviations occuring at intermediate values of the mass ratio, as expected, since this is the region where $g_c$ reaches its minimum and consequently the strong-coupling condition is less respected.

\section{Concluding Remarks}\label{conclusions}
In this paper, we have studied a resonant Bose-Fermi mixture in the presence of a mass and/or density imbalance. We have first analyzed the finite temperature phase diagram for the specific case of a $^{87}$Rb $-^{40}$K mixture. We have found that the overall shape of the phase diagram is similar to what found previously for equal masses. The critical temperature starts from the noninteracting value ($T_0= 3.31 n_{\rm B}^{2/3}/m_{\rm B}$) in the weak-coupling limit and eventually vanishes at a critical coupling $g_c$,  when the effect of the boson-fermion coupling is so strong that the Bose-Einstein condensate is completely depleted.
The critical temperature presents a weak maximum at intermediate coupling, not found for equal masses.  We have explained this feature as resulting from the balance between the hardening of the boson dispersion due to the boson-fermion interaction,  and the condensate depletion due to molecular correlation. The predominance of one effect over the other one depends on a fine tuning of the physical parameters, thus explaining the different behavior found for different cases.

We have then considered the zero-temperature limit. We have found that the critical coupling $g_c$ is significally affected by the mass imbalance, in particular for $m_{\rm B}/m_{\rm F}<1$. In this case, for all values of the density imbalance, we have obtained a rapid increase of $g_c$ as $m_{\rm B}/m_{\rm F}\to 0$.
We have also found that the density imbalance influences only weakly the critical coupling, as previously found for equal masses~\cite{Fra10}. On the other hand, a sufficiently large density imbalance produces quite a remarkable effect on the 
bosonic momentum distribution function. We have found indeed that this function  is completely depleted for momenta with magnitude below a certain value 
$q_0$. This value is set by the difference between the Fermi momenta associated with the unpaired fermions and the composite-fermions, respectively. Such a momentum distribution function reflects the constraints on the internal molecular wave-function imposed by the presence of a large Fermi surface associated with the unpaired fermions. In this respect, the physics underlying this phenomenon is similar to that of the ``Sarma phase'' for a highly-polarized Fermi gas in the strong-coupling limit of the BCS-BEC crossover~\cite{Sar63,pie06,She07,Rad10,Che10,sub10}. In particular, the bosonic momentum distribution function is similar to the momentum distribution function of the minority species in a polarized Fermi gas. Interestingly, the fact that in the present case the molecules cannot condense, but rather fill a Fermi sphere, 
smears some features of the momentum distribution function, but still preserves the presence of an empty region in the momentum distribution function. 
Such an interesting effect  could be measured in a Bose-Fermi mixture in the molecular limit by using the same technique used to measure the momentum distribution of an ultracold Fermi gas in the BCS-BEC crossover~\cite{reg05}.

\acknowledgments
Partial support by the Italian MUR under Contract Cofin-2009 ``Gas quantistici fuori equilibrio'' is acknowledged.

\appendix
\section{The one boson limit}
In this Appendix we consider the ``one boson'' limit $(n_{\rm F}-n_{\rm B})/(n_{\rm F}+n_{\rm B})\to 1$ of our equations. In this limit the fermions become free due to the vanishing boson density. One has then $\Sigma_{\rm F}= 0$ and $\mu_{\rm F}=(6\pi^2 n_{\rm F})^{2/3}/(2 m_{\rm F})= 2^{2/3} E_{\rm F}$. The boson self-energy $\Sigma_{\rm B}$ remains instead finite and is determined  by  Eq.~(\ref{selfbzero}), with no simplifications with respect to the case at finite boson density. 
The full knowledge of $\Sigma_{\rm B}$ is however not necessary to calculate  $\mu_{\rm B}$ in the limit $n_{\rm B}\to 0$. A study of the analytic structure of $\Sigma_{\rm B}$ in the complex frequency space along the lines of Ref.~\onlinecite{com09} shows that the limit $n_{\rm B}\to 0$ corresponds, for $g > g_c$, to the requirement that the minimum of the composite fermion dispersion $\Omega_0(\textbf{P})$ occurs exactly at zero frequency.  By introducing the retarded composite-fermion propagator via the replacement $\Gamma^{\rm R}({\bf P}, \omega)\equiv \Gamma({\bf P}, i\Omega \to \omega + i0^+)$, 
the dispersion  $\Omega_0(\textbf{P})$ is determined by the pole of $\Gamma^{\rm R}({\bf P}, \omega)$ in the complex plane.
By assuming the minimum of $\Omega_0(|\textbf{P}|)$  to occur at $\textbf{P}=0$, one has then the equation $\Gamma^{\rm R}(0,0)^{-1}=0$, which determines $\mu_{\rm B}$ at a given coupling and mass ratio. 
The critical coupling $g_c$, on the other hand, is determined by the equation $\mu_{\rm B}=\Sigma_{\rm B}(0,0)$, as for finite density.  In the limit $n_{\rm B}\to 0$ only the pole of the fermionic Green's function contributes to the frequency integral in Eq.~(\ref{selfbzero}), yielding:
\begin{equation}\label{selfbzerozero}
\Sigma_{\rm B}(0,0)=-\int\!\!\frac{d {\bf P}}{(2\pi)^{3}}\Theta[-\xi_{\rm F}(\textbf{P})]\Gamma^{\rm R}(\textbf{P}, \xi_{\rm F}(\textbf{P})),
\end{equation}
where
\begin{eqnarray}\label{gammaoneB}
\Gamma^{\rm R}(\textbf{P}, \xi_{\rm F}(\textbf{P}))&=&\left[ -\frac{m_{r}}{2 \pi a} \right.\nonumber\\
&+&\left. \frac{m_r^{3/2}}{\sqrt{2} \pi}\sqrt{\vert\mu_{\rm B}\vert- \frac{m_{\rm B}\textbf{P}^{2}}{2m_{\rm F}M}} + I_{\rm F}(\textbf{P})\right]^{-1}
\end{eqnarray}
and
\begin{equation}\label{Ifapp}
I_{\rm F}(\textbf{P})\equiv\int \frac{d {\bf p}}{(2\pi)^{3}} \frac{\Theta[-\xi_{\rm F}(\frac{m_{\rm F}}{M}\bf{P}-{\bf p})]}{\frac{{\bf p}^2}{2m_r}-\frac{m_{\rm B}}{m_{\rm F}} \frac{\textbf{P}^{2}}{2M}+ \vert\mu_{\rm B}\vert}.
\end{equation}
A calculation of the above integral yields
\begin{widetext}
\begin{equation}\label{Iffin}
I_{\rm F}(\textbf{P})=\frac{m_r  k_{\mu_{\rm F}}}{2\pi^2} + 
\frac{m_{\rm B}L_{{\bf P}}}{4 \pi^2} \frac{k_{\mu_{\rm F}}^2+2m_r \vert\mu_{\rm B}\vert -2(\frac{\bar{m}}{M}|\textbf{P}|)^2}{2|\textbf{P}|} -\frac{m_r R_{{\bf P}}}{2\pi^2}
\left[\arctan{\left(\frac{ k_{\mu_{\rm F}}+ \frac{m_{\rm F}}{M}|\textbf{P}|}{R_{{\bf P}}}\right)}+\arctan{\left(\frac{k_{\mu_{\rm F}}-\frac{m_{\rm F}}{M}|\textbf{P}|}{R_{{\bf P}}}\right)}\right],
\end{equation}
\end{widetext}
 where $R_{{\bf P}}\equiv[2 m_r \vert \mu_{\rm B} \vert - \left(|\textbf{P}| m_{\rm B}/{M}\right)^2]^{1/2}$, while 
\begin{equation}\label{LR}
L_{{\bf P}}\equiv\ln\left[\frac{(k_{\mu_{\rm F}}+|\textbf{P}|)(k_{\mu_{\rm F}}+\frac{\delta m}{M}|\textbf{P}|)+2 m_r \vert\mu_{\rm B}\vert}{(k_{\mu_{\rm F}}-|\textbf{P}|)(k_{\mu_{\rm F}}-\frac{\delta m}{M}|\textbf{P}|)+2 m_r \vert\mu_{\rm B}\vert}\right],
\end{equation}
where $\delta{m}=m_{\rm F}-m_{\rm B}$ and $\bar{m}^2=(m_{\rm F}^2+m_{\rm B}^2)/2$.
By using the expressions (\ref{gammaoneB}) to (\ref{LR}), the self-energy $\Sigma_{\rm B}(0,0)$ can then be calculated easily by a simple one-dimensional 
integral over $|{\bf P}|$. At $g_c$ the equation $\mu_{\rm B}=\Sigma_{\rm B}(0,0)$ then yields:
\begin{equation}
\mu_{\rm B}=-\int_0^{k_{\mu_{\rm F}}}\frac{d |{\bf P}|}{2\pi^2}\Gamma^{\rm R}(|\textbf{P}|, \xi_{\rm F}(|\textbf{P}|)).
\label{mub}
\end{equation}

The equation $0=\Gamma^{\rm R}(0,0)^{-1}$ yields finally
\begin{widetext}
\begin{equation}
0 = -\frac{m_r}{2\pi a}+\frac{m_{r}^{3/2}}{\pi}\sqrt{\vert \mu\vert}
+\frac{m_r}{\pi^2}\left[k_{\mu_{\rm F}}-2 \sqrt{m_r \vert \mu \vert}\arctan{\left(\frac{k_{\mu_{\rm F}}}{2\sqrt{m_r \vert \mu \vert}}\right)}\right]
\label{pole}
\end{equation}
\end{widetext}
The simultaneous solution of Eqs.~(\ref{mub}) and (\ref{pole}) allows to obtain
$g_c$ and $\mu_{\rm B}$ for a given mass ratio. Note finally that, even though we have derived Eqs.~(\ref{mub}) and (\ref{pole}) as a limiting case of our equations for a Bose-Fermi mixture, they describe also the polaron-to-molecule transition in a Fermi-Fermi mixture, since in this limit the statistics of the minority species becomes immaterial. We have verified, indeed, that our results for $g_c$ in this limit agree with the results reported in Ref.~\cite{com09} for the ``first-level approximation''.

\subsection{Asymptotic expressions for $g_c$ at small and large mass ratios}

We conclude this appendix by presenting the derivation of the asymptotic expressions (\ref{asysmall}) and (\ref{asylarge1}). We assume $\mu_{\rm B}$ to be large and negative as it occurs when $g$ is large. The validity of this assumption is verified by the asymptotic expressions for $g_c$ that are obtained  accordingly. The chemical potential $\mu=(\mu_{\rm B}+\mu_{\rm F})/2$ is then also large and negative. By expanding Eq.~(\ref{pole}) in powers of $k_{\mu_{\rm F}}/\sqrt{m_r \vert \mu \vert}$ one obtains 
\begin{equation}
2\mu\approx-\epsilon_0 + \frac{2}{3\pi m_r} k_{\mu_{\rm F}}^3 a .
\label{mu}
\end{equation}
or, equivalently
\begin{equation}
\mu_{\rm B}\approx-\epsilon_0 -\mu_{\rm F}+ \frac{2}{3\pi m_r} k_{\mu_{\rm F}}^3 a \label{muBapp}.
\end{equation}

Note that the subleading term $\frac{2}{3\pi m_r} k_{\mu_{\rm F}}^3 a$ in Eq.~(\ref{mu}) describes the mean-field repulsion between the molecule forming in the strong-coupling limit and the fermions, as it can be seen by casting it in the form $n_{\rm F} \frac{2\pi}{m_{\rm MF}} a_{\rm MF}$, where $m_{\rm MF}= M m_{\rm F}/(m_{\rm F}+M)$ is the reduced mass of a molecule and one fermion, while
\begin{equation} 
a_{\rm MF}=\frac{(1 + m_{\rm F}/m_{\rm B})^2}{1/2+ m_{\rm F}/m_{\rm B}} a
\end{equation}
is the molecule-fermion scattering length within the Born approximation \cite{isk07}.

\begin{figure}[t]
\epsfxsize=8cm
\epsfbox{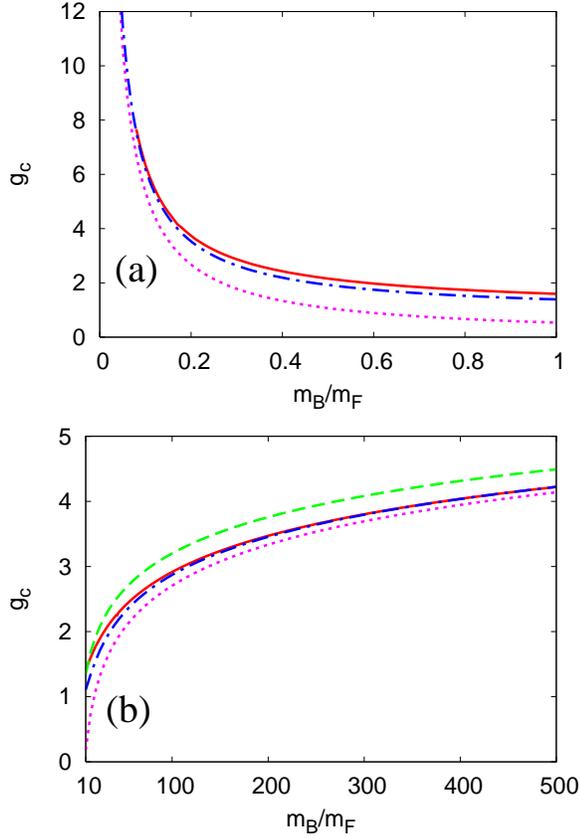}
\caption{(Color online) (a) The asymptotic expression (\ref{a3}) for $g_c$ at small mass ratios (dashed-dotted line) is compared with the full numerical solution (full line). The dotted curve is obtained by neglecting the subleading term within the brackets in (\ref{a3}).
(b) The asymptotic expression (\ref{a4}) for $g_c$ at large mass ratios (dashed-dotted line) is compared with the full numerical solution (full line). The dashed curve is obtained by neglecting the second and third terms on the right-hand side of (\ref{a4}), while the dotted curve neglects only the third term.} 
\label{asymptotic}
\end{figure}

The large and negative value of $\mu_{\rm B}$ then implies $I_{\rm F}(\textbf{P})\approx n_{\rm F}/\vert\mu_{\rm B}\vert\approx n_{\rm F}/\epsilon_0$, as it can be seen more easily directly from Eq.~(\ref{Ifapp}).
 The expansion of (\ref{gammaoneB}) for  $|\mu_{\rm B}|$ large then yields
\begin{equation}
\Gamma^{\rm R}(\textbf{P}, \xi_{\rm F}(\textbf{P}))\approx \frac{2\pi}{m_r^2 a}\frac{1}{\mu_{\rm F}-\alpha \textbf{P}^2}
\end{equation}
where  we have used Eq.~(\ref{mu}), and $\alpha\equiv m_{\rm B}/(2 M m_{\rm F})$. By using Eq.~(\ref{selfbzerozero}) we obtain
\begin{eqnarray}
\Sigma_{\rm B}(0,0) \approx\frac{1}{\alpha \pi m_r^2 a }\left[ k_{\mu_{\rm F}} +
\frac{\sqrt{\mu_{{\rm F}}/\alpha}}{2}\ln\frac{\sqrt{\mu_{{\rm F}}/\alpha} - k_{\mu_{\rm F}}}{k_{\mu_{\rm F}} + \sqrt{\mu_{{\rm F}}/\alpha}}\right]\nonumber\\
=\frac{k_{\mu_{\rm F}}}{\alpha \pi m_r^2 a }\left[1+\frac{\sqrt{1+\frac{m_{\rm F}}{m_{\rm B}}}}{2} \ln\frac{\sqrt{1+\frac{m_{\rm F}}{m_{\rm B}}}-1}{\sqrt{1+\frac{m_{\rm F}}{m_{\rm B}}}+1}\right]\phantom{aaaaaa}\label{sigmab1}
\end{eqnarray}

\noindent
By expanding the expression (\ref{sigmab1}) for a {\em small} mass ratio $m_{\rm B}/m_{\rm F}\ll 1$,  we obtain
\begin{eqnarray}
\Sigma_{\rm B}(0,0)&\approx& \frac{k_{\mu_{\rm F}}}{\alpha \pi m_r^2 a }\left[-\frac{1}{3}\frac{m_{\rm B}}{m_{\rm F}}+\frac{2}{15}\left(\frac{m_{\rm B}}{m_{\rm F}}\right)^2\right]\nonumber\\
&\approx&-\frac{2^{4/3}}{3\pi} \frac{k_{\rm F}}{a m_{\rm B}}\frac{m_{\rm F}}{m_{\rm B}}\left(1+ \frac{8}{5}\frac{m_{\rm B}}{m_{\rm F}}\right)
\end{eqnarray}
\noindent
The equation $\mu_{\rm B}=\Sigma_{\rm B}(0,0)$ then yields
\begin{equation}
\epsilon_0=\frac{2^{4/3}}{3\pi} \frac{k_{\rm F}}{a m_r}\frac{m_{\rm F}}{m_{\rm B}}\left(1+ \frac{8}{5}\frac{m_{\rm B}}{m_{\rm F}}\right),
\end{equation}
where we have kept only the leading term in the expression (\ref{muBapp}) for $\mu_{\rm B}$ (the term $\mu_{\rm F}$ would give a correction of order $(m_{\rm B}/m_{\rm F})^2$ to the asymptotic expression for $g_c$, with the last term in (\ref{muBapp}) contributing an even smaller correction). By substituting $\epsilon_0=1/(2 m_r a^2)$  we obtain then
\begin{equation}
\frac{1}{k_F a}=\frac{4 \cdot 2^{1/3}}{3\pi}\frac{m_{\rm F}}{m_{\rm B}}\left(1+ \frac{8}{5}\frac{m_{\rm B}}{m_{\rm F}}\right)\label{a3},
\end{equation}
which coincides with the expression (\ref{asysmall}) reported in Sec.~IV, and gives a large value of $g_c$ for small mass ratios, consistently with our starting assumption. The asymptotic expression (\ref{a3}) is compared with the full numerical calculation of $g_c$ in Fig.~\ref{asymptotic} (a).

The expansion of Eq.~(\ref{sigmab1}) for a  {\em large} mass ratio $m_{\rm B}/m_{\rm F}\gg 1$ yields instead 
\begin{eqnarray}
\Sigma_{\rm B}(0,0)&\approx&\frac{k_{\mu_{\rm F}}}{\alpha \pi m_r^2 a }\left( 1 + \frac{1}{2}\ln \frac{m_{\rm F}}{4 m_{\rm B}}\right)\\
&\approx&\frac{2^{4/3}}{\pi}\frac{k_{\rm F}}{a m_{\rm F}}\left( 1 + \frac{1}{2}\ln \frac{m_{\rm F}}{4 m_{\rm B}}\right)\label{sigmaBlarge}
\end{eqnarray}
where we have disregarded corrections to (\ref{sigmaBlarge}) smaller at least by a factor $m_{\rm F}/m_{\rm B}$. By equating (\ref{sigmaBlarge}) to (\ref{muBapp}) we get then
\begin{equation}
\epsilon_0 +\mu_{\rm F} - \frac{2  k_{\mu_{\rm F}}^3 a}{3\pi m_r} = \frac{2^{\frac{4}{3}}}{\pi}\frac{k_{\rm F}}{a m_{\rm F}}\left( -1 + \frac{1}{2}\ln \frac{4 m_{\rm B}}{m_{\rm F}}\right)
\end{equation}
yielding
\begin{equation}
\frac{1}{2 a^2 m_{\rm F}}= \frac{2^{\frac{1}{3}}k_{\rm F}}{\pi a m_{\rm F}}\!\!\left(\!\ln \frac{4 m_{\rm B}}{m_{\rm F}} - 2\!\right) - \frac{ 2^{\frac{2}{3}}k_F^2}{2 m_{\rm F}}  + \frac{4  k_{{\rm F}}^3 a}{3\pi m_{\rm F}}\label{a1}
\end{equation}
from which, by multiplying both sides of Eq.~(\ref{a1}) by $2 a m_{\rm F}/ k_{\rm F}$, we obtain
\begin{equation}
\frac{1}{k_{\rm F} a}  = \frac{2^{\frac{4}{3}}}{\pi}\!\left(\ln \frac{4 m_{\rm B}}{m_{\rm F}} -2 \right) - 2^{\frac{2}{3}}k_{\rm F} a  + \frac{8  (k_{{\rm F}}a)^2}{3\pi}\label{a2} 
\end{equation}
which shows that, to leading order, $g_c\approx \frac{2^{\frac{4}{3}}}{\pi} \ln \frac{m_{\rm B}}{m_{\rm F}}$ at large mass ratios. The slow log-dependence of $g_c$ on $m_{\rm B}/m_{\rm F}$ makes however significant to keep also a few subleading corrections to the leading behavior. The solution of Eq.~(\ref{a2}) by iteration yields then
\begin{eqnarray}
\frac{1}{k_{\rm F} a}  &=& \frac{2^{\frac{4}{3}}}{\pi}\left( \ln \frac{4 m_{\rm B}}{m_{\rm F}} -2 \right) - \frac{2^{\frac{2}{3}}}{\frac{2^{\frac{4}{3}}}{\pi}\left( \ln \frac{4 m_{\rm B}}{m_{\rm F}} -2 \right)}\nonumber 
\\
&+& \frac{8}{3\pi}\frac{1}{\left[\frac{2^{\frac{4}{3}}}{\pi}\left( \ln \frac{4 m_{\rm B}}{m_{\rm F}} -2 \right)\right]^2}\label{a4}
\end{eqnarray}
 which ignores corrections of order 
$1/(\ln \frac{m_{\rm B}}{m_{\rm F}})^3$ or higher, and coincides with the expression (\ref{asylarge1}) reported in Sec.~IV.  The asymptotic expression (\ref{a4}) is compared with the full numerical calculation of $g_c$ in Fig.~\ref{asymptotic} (b).



\begin{thebibliography}{99}

\bibitem{Viv00}
L. Viverit, C.J. Pethick, and H. Smith, 
Phys. Rev. A \textbf{61}, 053605 (2000).
\bibitem{Yi01}
X. X. Yi and C. P. Sun, 
Phys. Rev. A \textbf{64}, 043608 (2001).
\bibitem{Alb02} A. P. Albus, S. A. Gardiner, F. Illuminati, and M. Wilkens,
Phys. Rev. A \textbf{65}, 053607 (2002).
\bibitem{Viv02} L. Viverit and S. Giorgini,
Phys. Rev. A {\bf 66}, 063604 (2002).
\bibitem{Rot02}
R. Roth and H. Feldmeier, 
Phys. Rev. A \textbf{65}, 021603(R) (2002).
\bibitem{Li03}
X.-J. Liu, M. Modugno, and H. Hu, 
Phys. Rev. A \textbf{68}, 053605 (2003).
\bibitem{Lew04}
M. Lewenstein, L. Santos, M. A. Baranov, and H. Fehrmann, 
Phys. Rev. Lett. \textbf{92}, 050401 (2004).

\bibitem{Mod02}
G. Modugno, G. Roati, F. Riboli, F. Ferlaino, R. J. Brecha, and M. Inguscio,
Science \textbf{297}, 2240 (2002).

\bibitem{Pow05}
S. Powell, S. Sachdev, and H. P. Buchler,
Phys. Rev. B \textbf{72}, 024534 (2005).
\bibitem{Dic05}
D. B. M. Dickerscheid, D. van Oosten, E. J. Tillema, and H. T. C. Stoof, 
Phys. Rev. Lett. \textbf{94}, 230404 (2005).
\bibitem{Avd06}
A. V. Avdeenkov, D. C. E. Bortolotti, and J. L. Bohn,
Phys. Rev. A \textbf{74}, 012709 (2006).
\bibitem{Bor08}
D. C. E. Bortolotti, A. V. Avdeenkov, and J. L. Bohn,
 Phys. Rev. A \textbf{78}, 063612 (2008).


\bibitem{Mar08}
F. M. Marchetti, C. J. M. Mathy, D. A. Huse, and M.M. Parish,
 Phys. Rev. B \textbf{78}, 134517 (2008).

\bibitem{Gun06}
K. G\"unter, T. Stoferle, H. Moritz, M. Kohl, and T. Esslinger,
Phys. Rev. Lett. {\bf 96}, 180402 (2006).


\bibitem{Osp06}
C. Ospelkaus, S. Ospelkaus, L. Humbert, P. Ernst, K. Sengstock, and K. Bongs,
Phys. Rev. Lett. {\bf 97}, 120402 (2006).


\bibitem{Osp06b}
S. Ospelkaus, C. Ospelkaus, L. Humbert, K. Sengstock, and K. Bongs,
Phys. Rev. Lett. {\bf 97}, 120403 (2006).

\bibitem{Zir08}
J. J. Zirbel, K.-K. Ni, S. Ospelkaus, J. P. D'Incao, C. E. Wieman, J. Ye, and D. S. Jin,
Phys. Rev. Lett. {\bf 100}, 143201 (2008).

\bibitem{Ni08}
K.-K. Ni, S. Ospelkaus, M. H. G. de Miranda, A. Pe'er, B. Neyenhuis, J. J. Zirbel, S. Kotochigova, P. S. Julienne, D. S. Jin, and J. Ye,
Science {\bf 322}, 231 (2008).

\bibitem{Wu11}
C.-H. Wu, I. Santiago, J. W. Park, P. Ahmadi, and M. W. Zwierlein, 
Phys. Rev. A {\bf 84}, 011601 (2011). 

\bibitem{Park11}
J. W. Park, C.-H. Wu, I. Santiago,  T. G. Tiecke, S. Will, P. Ahmadi, and M. W. Zwierlein,
 Phys. Rev. A {\bf 85}, 051602 (2012).

\bibitem{Heo12}
M.-S. Heo, T. T. Wang, C. A. Christensen, T. M. Rvachov, D. A. Cotta, J.-H. Choi, Y.-R. Lee, W. Ketterle,
arXiv:1205.5304 (2012).

\bibitem{Sim05}
S. Simonucci, P. Pieri, and G.C. Strinati, Europhys. Lett. {\bf 69}, 713 (2005). 



\bibitem{Kag04}
M.Y. Kagan,  I. V. Brodsky, D. V. Efremov, and A. V. Klaptsov, 
Phys. Rev. A \textbf{70}, 023607 (2004).

\bibitem{Bar08}
X. Barillier-Pertuisel, S. Pittel, L. Pollet, and P. Schuck,
Phys. Rev. A  \textbf{77}, 012115 (2008).

\bibitem{Pol08}
L. Pollet, C. Kollath, U. Schollw\"ock, and M. Troyer,
Phys. Rev. A \textbf{77}, 023608 (2008).

\bibitem{Riz08}
M. Rizzi and A. Imambekov,
Phys. Rev. A \textbf{77}, 023621 (2008).

\bibitem{Tit09}
I. Titvinidze, M. Snoek, and W. Hofstetter, 
Phys. Rev. B {\bf 79}, 144506 (2009).

\bibitem{Mar09}
F. M. Marchetti, T. Jolicoeur, and M. M. Parish,
 Phys. Rev. Lett. {\bf 103}, 105304 (2009).

\bibitem{Sto05}
A. Storozhenko, P. Schuck, T. Suzuki, H. Yabu, and J. Dukelsky,
 Phys. Rev. A \textbf{71}, 063617 (2005).


\bibitem{Wat08}
T. Watanabe, T. Suzuki, and P. Schuck,
Phys. Rev. A \textbf{78}, 033601 (2008).


\bibitem{Fra10}
E. Fratini and P. Pieri, 
Phys. Rev. A \textbf{81}, 051605(R) (2010).

\bibitem{Lud11}
D. Ludwig, S. Floerchinger, S. Moroz, and C. Wetterich, 
Phys. Rev. A {\bf 84}, 033629 (2011).

\bibitem{Pie00} P. Pieri and G.C. Strinati, Phys. Rev. B {\bf 61}, 15370 (2000).
\bibitem{Yu11}
Z.-Q Yu, S. Zhang and H. Zhai, 
Phys. Rev. A  {\bf 83}, 041603(R) (2011).

\bibitem{popov} See, e.g., chap.~6 of V. N. Popov, {\em Functional Integrals and Collective Excitations}, Cambridge Univ.~Press, Cambridge (1987). 

\bibitem{footnote-zeroT} See, e.g., Sec.~29 of A. Fetter and J. D.Walecka,  {\em Quantum theory of many-particle systems}, Mc-Graw Hill, New York (1971) and J. M. Luttinger and J. C. Ward, Phys. Rev. {\bf 118}, 1417 (1960). Problems in the replacement of the frequency sums with integrals, in the zero-temperature limit, normally arise in the presence of double (or higher) poles of the summands in the complex frequency plane. We have not met these type of singularities in our case. In any case, the good matching of the finite temperature results with those obtained at zero temperature verify {\em a posteriori} the validity of the procedure.
 
\bibitem{Holz03}
M. Holzmann, J. N. Fuchs, G. Baym, J. P. Blaizot, F. Lalo{\"e},
C. R. Physique {\bf 5}, 21 (2004).

\bibitem{pro08} N. V. Prokof'ev and B. V. Svistunov, Phys. Rev. B {\bf 77}, 125101 (2008).

\bibitem{vei08} M. Veillette {\em et al.},
Phys. Rev. A {\bf 78}, 033614 (2008).
\bibitem{mas08}
P. Massignan, G. M. Bruun, and H. T. C. Stoof, 
Phys. Rev. A  {\bf 78}, 031602(R) (2008).
\bibitem{mor09}
C. Mora and F. Chevy,
Phys. Rev. A {\bf 80}, 033607 (2009).
\bibitem{pun09}
M. Punk, P. T. Dumitrescu, and W. Zwerger,
Phys. Rev. A {\bf 80}, 053605 (2009).
\bibitem{com09}
R.~Combescot, S.~Giraud, and X.~Leyronas,
Europhys. Lett {\bf 88}, 60007 (2009).

\bibitem{sch09} A. Schirotzek, , C.-H. Wu, A. Sommer, and M. W. Zwierlein,
Phys. Rev. Lett. {\bf 102}, 230402 (2009).

\bibitem{Mat11} C. J. M. Mathy, M. M. Parish, and D. A. Huse, 
Phys. Rev. Lett. {\bf 106}, 166404 (2011)
\bibitem{Song11} 
J.-L. Song and F. Zhou, 
Phys. Rev. A {\bf 84}, 013601 (2011).
\bibitem{braaten06} E. Braaten, H. W. Hammer, Phys. Rep. {\bf 428}, 259 (2006).
\bibitem{cor12} 
M. Correggi, G. Dell'Antonio, D. Finco, A. Michelangeli, and A. Teta, arXiv:1201.5740 (2012).
\bibitem{Kar07} O. I. Kartavtsev and A. V. Malykh, J. Phys. B {\bf 40}, 1429 (2007).
\bibitem{par11} M. M. Parish, Phys. Rev. A {\bf 83}, 051603 (2011).
\bibitem{chevy}  F. Chevy, Phys. Rev. A {\bf 74}, 063628 (2006).
\bibitem{tan} S. Tan, Ann. Phys. {\bf 323}, 2952 (2008); {\em ibidem}, 2971 (2008).
\bibitem{Sar63} G. Sarma,  J. Phys. Chem. Solids {\bf 24}, 1029 (1963).
\bibitem{pie06} P. Pieri and G.C. Strinati,
Phys. Rev. Lett. {\bf 96}, 150404 (2006).
\bibitem{She07}
 D. E. Sheehy and L. Radzihovsky, Ann. Phys. {\bf 322}, 1790  (2007).
\bibitem{Rad10} L. Radzihovsky and D. E. Sheehy,  Rep. Prog. Phys. {\bf 73}, 076501 (2010).
\bibitem{Che10} F. Chevy and C. Mora, Rep. Prog. Phys. {\bf 73}, 112401 (2010).
\bibitem{sub10} P. Pieri, D. Neilson, and G.C. Strinati, Phys. Rev. B {\bf 75}, 113301 (2007); A. L. Subasi, P. Pieri, G. Senatore, and B. Tanatar,
Phys. Rev. B {\bf 81}, 075436 (2010).

\bibitem{reg05} C. A. Regal, M. Greiner, S. Giorgini, M. Holland, and D. S. 
Jin,  Phys. Rev. Lett. {\bf 95}, 250404 (2005).
\bibitem{isk07} M. Iskin and C. A. R. Sa de Melo, Phys. Rev. A {\bf 76}, 013601 (2007)
\end{thebibliography}
\end{document}